\documentclass{article}

\usepackage{arxiv}

\usepackage[utf8]{inputenc} 
\usepackage[T1]{fontenc}    
\usepackage{hyperref}       
\usepackage{url}            
\usepackage{booktabs}       
\usepackage{amsfonts}       
\usepackage{nicefrac}       
\usepackage{microtype}      
\usepackage{lipsum}
\usepackage{graphicx}
\graphicspath{ {./images/} }
\usepackage{framed,multirow}

\usepackage{amssymb}
\usepackage{latexsym}

\usepackage{url}
\usepackage{xcolor}
\usepackage{todonotes}
\usepackage{multicol}
\usepackage{longtable}
\usepackage{supertabular}
\usepackage{cuted}
\usepackage{amsmath}
\usepackage[ruled,vlined]{algorithm2e}
\usepackage[normalem]{ulem}

\useunder{\uline}{\ul}{}

\DeclareUnicodeCharacter{2212}{-}

\definecolor{mycolor}{HTML}{000000}

\newcommand*{\equationref}[1]{%
  Eq. \eqref{#1}}
\newcommand*{\tableref}[1]{%
  Table \ref{#1}}
\newcommand*{\figureref}[1]{%
  Fig. \ref{#1}}

\title{Explainable Image Quality Assessment for Medical Imaging}

\author{
 Caner Ozer \\
  Computer Engineering Department\\
  Istanbul Technical University\\
  Istanbul, Turkey \\
  \texttt{ozerc@itu.edu.tr} \\
   \And
 Arda Guler \\
  Mehmet Akif Ersoy \\ Thoracic and Cardiovascular Surgery \\ Training and Research Hospital \\
  Istanbul, Turkey \\
  \And
 Aysel Turkvatan Cansever \\
  Mehmet Akif Ersoy \\ Thoracic and Cardiovascular Surgery \\ Training and Research Hospital \\
  Istanbul, Turkey \\
  \And
 Ilkay Oksuz \\
  Computer Engineering Department\\
  Istanbul Technical University\\
  Istanbul, Turkey \\
}

\begin{document}
\maketitle
\begin{abstract}
Medical image quality assessment is an important aspect of image acquisition, as poor-quality images may lead to misdiagnosis. Manual labelling of image quality is a tedious task for population studies and can lead to misleading results. While much research has been done on automated analysis of image quality to address this issue, relatively little work has been done to explain the methodologies. In this work, we propose an explainable image quality assessment system and validate our idea on two different objectives which are foreign object detection on Chest X-Rays (Object-CXR) and Left Ventricular Outflow Tract (LVOT) detection on Cardiac Magnetic Resonance (CMR) volumes. We apply a variety of techniques to measure the faithfulness of the saliency detectors, and our explainable pipeline relies on NormGrad, an algorithm which can efficiently localise image quality issues with saliency maps of the classifier. We compare NormGrad with a range of saliency detection methods and illustrate its superior performance as a result of applying these methodologies for measuring the faithfulness of the saliency detectors. We see that NormGrad has significant gains over other saliency detectors by reaching a repeated Pointing Game score of $0.853$ for Object-CXR and $0.611$ for LVOT datasets. \end{abstract}


\section{Introduction}\label{intro_main}
Maintaining a high image quality during a medical scan is essential for obtaining a clear diagnosis of a patient. However, some distortions in the image quality may hinder an accurate analysis of the acquired medical scan, and this would eventually lead to inaccurate diagnostic conclusions by the physician. The reasons do include physiological and patient-based motion artefacts in Magnetic Resonance Imaging (MRI) \cite{Ma2020, Oksuz2020}, mistriggering-based motion artefacts in MRI, which happen due to incorrect ECG triggering during the MRI scan \cite{Oksuz2019}, or the contrast problem in CT due to low-dose imaging \cite{Goldman2007}. Assuming the existence of a reference image, it is possible to evaluate the level of distortion through image quality assessment metrics, such as Peak Signal-to-Noise Ratio (PSNR), Structural Similarity Index \cite{Wang2003}, or among more recent methods: HaarPSI \cite{REISENHOFER2018} and DISTS \cite{Ding2020}. However, often such reference does not exist and  manual labelling based on qualitative analysis is required \cite{Yoneyama2012}. This procedure is generally tedious and prone to human error.

Obtaining a reference image is not possible for some types of medical image quality assessment problems. Some of the image quality problems would arise from foreign object appearance, such that some undesirable objects would appear in a medical scan. This case can be frequently encountered in localised image quality assessment problem, where only a small proportion of the medical scan can cause the degradation of image quality. For instance during a Chest X-Ray scan, some foreign objects such as clips and buttons may appear on the lungs or around the cardiac region, which would prevent the visibility for an accurate diagnosis \cite{CXR20}. In addition to that, it is also possible to encounter some undesirable regions, such as Left Ventricular Outflow Tract (LVOT) appearance in the Cardiac MRIs with long axis view \cite{Oksuz2018}. This is generally caused by poor planning of the cardiac scan, which results in an off-axis view. Since we cannot evaluate the level of quality directly through an image quality assessment metric, we rely on automatic image quality assessment mechanisms to assess the quality of a medical scan, with an automated machine learning algorithm. This algorithm's output can be used to filter low-quality images in large population studies or to increase the quality of the medical scans with additional quality improvement pipelines. Determining image quality successfully is also an essential task before developing machine learning frameworks for down-stream tasks (e.g. segmentation, disease diagnosis).

Explainability has become a crucial aspect of evaluating the trustability of automatic diagnostic systems in particular in black-box deep learning applications \cite{VANDERVELDEN2022}. Given that a deep classifier provides us only the predicted label for a given medical scan, it is imperative to find out the cues which can help to make sure that the decision has been made based on the relatable sets of features. Also, it has recently been show that physicians, who lack task-specific knowledge, can benefit from the recommendations of explainable AI more than human advices in the way of visual annotations \cite{Gaube2023}. However, an explainability analysis is not well-studied for understanding the deep learning models for medical image quality assessment, even though this research field has recently being challenged in \cite{Arun2021} and \cite{Jin2022}. Also, the methodology that is responsible for the explainability analysis should be quantitatively analysing the precision and faithfulness \cite{Adebayo2018} aspects of the saliency maps. These two aspects become vital especially in the presence of abnormalities with the medical scan in terms of the medical image quality.

In this paper, we propose a consistent explainable image quality assessment pipeline based on NormGrad \cite{Rebuffi20}. This paper has been built upon our previous work, \cite{Ozer2021}, in which we proposed to use NormGrad to localise the regions that cause the image quality degradation on Chest X-Rays by pointing the foreign objects. Here, we extend this approach on 4-chamber Cardiac MRI scans by localizing the LVOT regions and show the strength of NormGrad by performing supplementary experiments on a number of saliency detectors. To the best of our knowledge, this is the first approach that uses NormGrad for medical image quality assessment while being the first paper to perform a comprehensive explainable image quality assessment on Chest X-Ray and Cardiac MRI data. The main contributions of our approach are listed as follows:

\begin{itemize}
    \item NormGrad is proposed to detect the target regions of interest which indicate the image quality problems.
    \item Extensive validation experiments are performed to assess the performance of NormGrad, by contrasting other explainability methods. We measured the effect of using smoothing on the saliency maps, randomizing the neural network models, repeating the experiments from scratch a different number of times, changing the network architecture.
    \item Difference of Means (DoM) metric is proposed to show the consistency of a saliency detector whenever we change the network architecture.
    \item The pipeline is validated by using Object-CXR (X-Ray) and LVOT Detection (Cardiac MRI) datasets.
\end{itemize}

The remainder of the paper is organized as follows. Section \ref{sec:related_work} provides the relevant literature to this work. Section \ref{sec:method} defines the mathematical background of the methods that are used for our explainability analysis, and Section \ref{sec:experiments} presents our results of qualitative and quantitative analyses to evaluate the performance of a variety of explainability methods. Further discussion is made on the methods in Section \ref{sec:diss_conc}, prior to providing our conclusions.

\section{Literature Review}\label{sec:related_work}

In this section, we provide an overview of the related works on image quality assessment, interpretability, and interpretable medical image quality assessment. 

\subsection{Image Quality Assessment}

Medical Image Quality Assessment (IQA) is becoming an emergent research field with the advent of deep learning-based approaches. It is crucial to maintain sufficient image quality for alleviating potential diagnostic errors made either automatic or manual image quality assessment procedures, which can be found in \cite{Chow2016} in detail. Predominantly, the domain knowledge holds the key while assessing the quality. Therefore, most of the approaches can only be applicable on a single modality and specific type of image quality assessment problem. For instance, \cite{Sujit2018} proposed a model for structural brain MRI quality assessment where they used a small CNN architecture for measuring the overall image quality automatically.  
Also, \cite{Hu2022} proposed a diagnostic quality assessment dataset composed of Chest X-Rays and used a semantic segmentation-guided multi-label classification framework. 
\cite{Mortamet2009} named several factors for the quality degradation on brain MRIs and proposed two different quality measures based on a voxel-wise quality assessment.
\cite{Abdi2017} treated the quality assessment as a regression problem and assessed the quality of echocardiography cines in five different views by using a convolutional neural network with different multi-head LSTM layers.
Similarly \cite{Ozer2021} assessed the quality of Chest X-Ray scans depending on the existence of foreign objects. \cite{Chen2021} constructed an intravascular ultrasound dataset consisting of 3 quality labels, e.g., low, middle and high, and used a ResNet-18 model \cite{He2016} with squeeze and excitation modules \cite{Hu2018} to predict the quality label. \cite{Ma2020} developed a T2-weighted abdomen MR dataset consisting of good quality and scans with motion artefacts, and used a 4-layer CNN framework to automatically assess the quality of the scans. \cite{Oksuz2020} further advanced in this motion artefact detection problem by not only adopting a synthetic motion artefact generation scheme, but also by performing detection and overcoming the quality issues during the reconstruction phase of the short-axis cardiac MRI scans. 

\subsection{Interpretability}
Interpretability is an important area of research at the intersection of medical image analysis and deep learning. Early approaches to interpretability in deep learning include the DeConvNet \cite{Zeiler2014} approach, which visualizes intermediate-level activations by reverting them to the original image dimensionality.  However, the use of gradient for the \textit{saliency map} visualizations has been initiated by Grad \cite{Simonyan2014}, which calculates the derivative of output with respect to the input image. Subsequently, due to the nature of the ReLU activation function and advancing network architectures, Guided Backpropagation \cite{Springenberg2015} was proposed to alter the gradient backpropagation mechanism by also considering the sign values of the upstream gradient. Following this, various methods that combine activations and gradients have become more prevalent in the general computer vision and medical image analysis literature, such as Input x Grad \cite{Shrikumar2016}, Grad-CAM \cite{Selvaraju2020}, Guided Grad-CAM \cite{Selvaraju2020}, and NormGrad \cite{Rebuffi20}. In parallel, Integrated Gradients \cite{Sundararajan2017}, Score-CAM \cite{Wang2020}, Smooth-Grad \cite{Smilkov2017}, and RISE \cite{Petsiuk2018} have proposed to use an iterative procedure to calculate the saliency map.

Recently, the use of saliency detectors to demonstrate diagnostic findings through heatmaps has become increasingly popular in medical imaging. For instance, \cite{Lin2022} used Grad-CAM to localise the brain regions that do and do not contribute to schizophrenia on the spatial source phase maps. \cite{Kim2022} modified the DenseNet architecture to improve the visualization of Grad-CAM without any fall in accuracy for arrhythmia detection using ECG. \cite{Park2022} adopted a Vision Transformer-based architecture for COVID-19 severity quantification and used a variety of CAM and deep taylor decomposition to show the saliency maps. \cite{Huang2022} leveraged the saliency maps to train a diabetic retinopathy grading framework in a self-supervised fashion.
Even though the research community demonstrates the findings shown by a saliency method qualitatively, there has yet to be a systematic analysis done to validate the relationship between a saliency method and the model's performance for the domain of medical image analysis. In this regard, we benchmark NormGrad with other saliency detection methods for the interpretable image quality assessment problem.

\section{Method} \label{sec:method}

In this section, we briefly describe the use of Input x Grad \cite{Shrikumar2016}, Guided Backpropagation \cite{Springenberg2015}, Grad-CAM \cite{Selvaraju2020}, Guided Grad-CAM and NormGrad \cite{Rebuffi20} frameworks for medical image quality assesment. These frameworks generally aid us in constructing the saliency maps of deep neural network-based classifiers. 

For all of these frameworks, we assume that there exists a pre-trained neural network such that we would like to extract the knowledge about some target layer $k_t$. We also define its preceding layers with $p$ and succeeding with $q$. Given an input image $\textbf{x} \in {\rm I\!R}^{C \times H \times W}$, where $C$ refers to the number of input channels and $H$ and $W$ correspond to the size of the image, we can also define the relation between $\textbf{x}^{in} \in {\rm I\!R}^{K \times H' \times W'}$, $\textbf{x}^{out} \in {\rm I\!R}^{K' \times H' \times W'}$, and the network output, $\textbf{y}$, as follows:

\begin{equation}
\begin{split}
\textbf{x}^{in} = p(\textbf{x})
 \\
 \textbf{x}^{out} = k_t(\textbf{x}^{in})
 \\
 \textbf{y} = q(\textbf{x}^{out}).
\end{split}
\end{equation}
We define the gradient w.r.t. the parameters of layer $k_t$ as $\textbf{g}^{out} \in {\rm I\!R}^{K' \times H' \times W'}$, the activations of the same layer as $\textbf{x}^{out}$, the gradient w.r.t. the input image $\textbf{x}$ as $\textbf{g}$. Unless stated otherwise, the gradient is calculated by assuming that $\textbf{y}$ is the ground-truth class label, while this may not be true for all samples.

\begin{figure*}
    \centering
    \includegraphics[width=0.9\columnwidth]{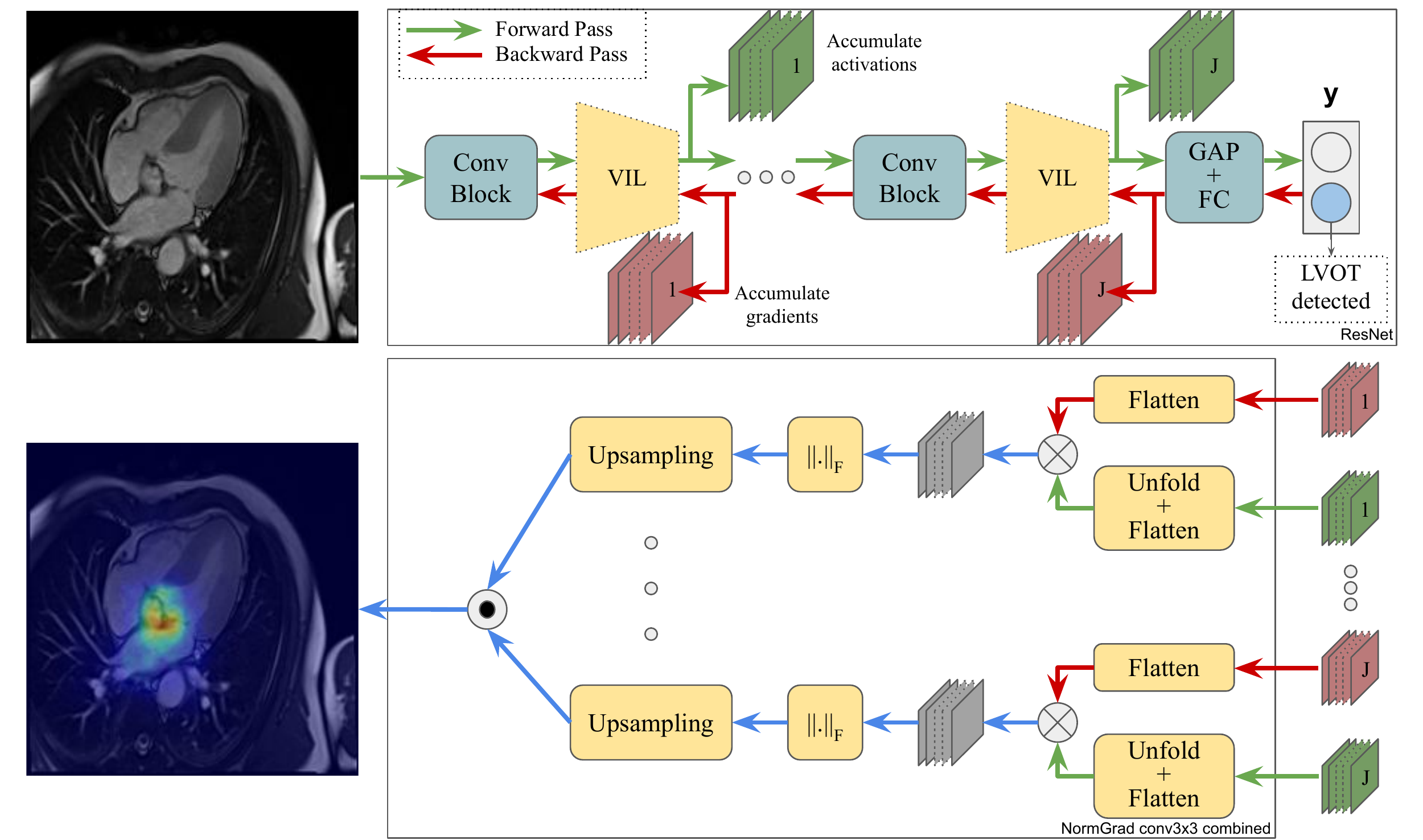}
    \caption{NormGrad Framework. The top part shows the flow of a neural network where multiple virtual identity layers (VILs) are placed right after the convolutional blocks to accumulate the activations and gradients of different locations. The bottom part demonstrates how they are used to obtain a unified heat map by using NormGrad. Red and blue colours within the heat maps point to regions with high and low values. Conv Block refers to a convolutional layer with batch normalisation and ReLU activation function, whereas GAP and FC operations stand for Global Average Pooling and Fully-Connected layers, respectively. Left Ventricular Outflow Tract appearance on a 4-chamber Cardiac MRI scan is shown on the input image with a red bounding box.}
    \label{fig:normgrad}
\end{figure*}

In addition, consider that a virtual identical layer $\Tilde{k}_t$, is present right after the layer $k_t$, whose output is $\Tilde{\textbf{x}}^{out}$ and satisfies the property $\Tilde{\textbf{x}}^{out} = \textbf{x}^{out}$. The purpose of adding this layer is to ensure that activations and gradients are collected from the same point of the network. Moreover, this layer can be chosen among bias, scaling, or convolutional layers. If the framework choice is NormGrad, we accumulate the output activations, $\Tilde{\textbf{x}}^{out}$, and the corresponding upstream gradient, $\textbf{g}^{out}$, of the layer $\Tilde{k}_t$. However, we omitted the results regarding the bias setting of NormGrad since it is unable to exploit the strength of activations, and the gradient is nothing different than a constant map whenever the virtual identity layer is placed at the end of the last convolutional layer.

\subsection{Input x Grad}
Input x Grad is one of the baseline methods for saliency detection, where the gradients of backpropagation are multiplied by the activation maps. In this way, it becomes possible to observe the peaks and troughs of this multiplication. Peaks demonstrate a correspondence between the input image and the gradient, whereas troughs highlight the differences between these two tensors. Formally, we perform an element-wise multiplication of $\textbf{x}$ with $\textbf{g}$ in \equationref{eq:ixg}, such that

\begin{equation}
    \textbf{m}_{IxG} = \textbf{x} \odot \textbf{g},
    \label{eq:ixg}
\end{equation}

where $\textbf{m}_{IxG}$ becomes the saliency map that is obtained by using Input x Grad.

\subsection{Guided Backpropagation}
In Guided Backpropagation, the gradient flow of a non-linearity is altered by performing the backward pass by using backpropagation and deconvolution approaches instantaneously. Considering that ReLU \cite{Glorot2011} is used as the choice of non-linearity, the upstream gradient is masked by using the forward-pass values of the ReLU activation. In this manner, it becomes possible to filter by considering the signs of the forward-pass during backpropagation. In the deconvolution approach, however, masking is performed by observing the backward-pass values instead of the forward-pass. In mathematical terms, for a ReLU activation in some layer $l_t$, we define backpropagation in \equationref{eq:bp}, deconvolution in \equationref{eq:deconv} and guided backpropagation in \equationref{eq:gbp},

\begin{equation}
    \textbf{g}^{pre}_{t} = (\textbf{h}_t > 0) \odot \textbf{g}_{t},
    \label{eq:bp}
\end{equation}

\begin{equation}
    \textbf{g}^{pre}_{t} = (\textbf{g}_{t} > 0) \odot \textbf{g}_{t},
    \label{eq:deconv}
\end{equation}

\begin{equation}
    \textbf{g}^{pre}_{t} = (\textbf{h}_t> 0) \odot (\textbf{g}_{t} > 0) \odot \textbf{g}_{t},
    \label{eq:gbp}
\end{equation}
where $\odot$ corresponds to the element-wise multiplication operation, and $\textbf{h}_t$ is the output of the ReLU activation function and $\textbf{g}_t$ is the upstream gradient that are present within the layer $l_t$. If we would like to calculate the saliency map by using guided backpropagation, we apply the \equationref{eq:gbp} instead of \equationref{eq:bp} for all of the ReLU layer gradient calculations within a neural network.

However, more recent neural network architectures, as in EfficientNet \cite{Tan2019}, tend to use the SiLU activation function \cite{Elfwing2018}, which stands for the Sigmoid Linear Unit, as shown in \equationref{eq:silu}. Here, $\sigma$ is the Sigmoid activation function which equals $\sigma(\textbf{x}) = \frac{1}{1 + e^{-\textbf{x}}}$, and $\textbf{x}$ is some input tensor. Taking the gradient w.r.t. $\textbf{x}$, we demonstrate the outcome in \equationref{eq:dsilu}.
\begin{equation}
    a_k(\textbf{x}) = \textbf{x} \odot \sigma(\textbf{x})
    \label{eq:silu}
\end{equation}

\begin{equation}
    \frac{d a_k(\textbf{x})}{d \textbf{x}} = \sigma(\textbf{x}) + \textbf{x} \odot \sigma(\textbf{x}) \odot (1 - \sigma(\textbf{x}))
    \label{eq:dsilu}
\end{equation}

Using the derivative of the SiLU activation function, we can also express the guided backpropagation when the SiLU activation function is used. In a similar fashion to \equationref{eq:gbp}, we can illustrate the guided backpropagation by showing the output of the SiLU activation function at layer $k_t$ with $\textbf{h}_t$ in \equationref{eq:gbp_silu}:
\begin{equation}
    \textbf{g}^{pre}_{t} = \frac{d a_k(\textbf{h}_t)}{d \textbf{h}_t} \odot \frac{d a_k(\textbf{g}_{t})}{d \textbf{g}_{t}} \odot \textbf{g}_{t}
    \label{eq:gbp_silu}
\end{equation}

For the saliency map calculation by guided backpropagation, we use the first convolutional layer within the network as the choice of $k_t$ and filter out all of the negative values of \textbf{g}. This becomes the saliency map, $\textbf{m}_{GBP}$, that has been obtained by using guided backpropagation.

\subsection{Grad-CAM}
There are four steps to construct the saliency maps after setting $k_t$ as the last convolutional layer of the neural network. First of all, a vector, $\alpha$, that corresponds to the importance weights for each filter is created for the last convolutional layer of a neural network by the relation,

\begin{equation}
    \alpha = \frac{1}{H \times W} \sum_{h, w} \textbf{g}^{out}[:, h, w]
\end{equation}

where $\textbf{g}^{out}$ is the gradient w.r.t. the parameters of the final convolutional layer of the neural network and $H \times W$ acts as a normalisation factor across all pixels. 
The $\alpha$ vector acts as an importance measure for each of the convolutional filters that are present within the layer. 
Then, a summation on the channel axis right after the element-wise multiplication takes place between the $\alpha$ and $\textbf{x}^{out}$ as in \equationref{eq:gc:sc}

\begin{equation}
    S = \sum_k \alpha_k \odot \textbf{x}_k^{out}
    \label{eq:gc:sc}
\end{equation}

which will result in the aggregated spatial contribution, $S$, or the saliency map. In addition, a further filtering procedure is applied to weaken the existence of depleted regions having negative values. For this reason, the ReLU activation function is adopted on $S$ to highlight the regions with non-negative values, as shown in \equationref{eq:gc:relu}.

\begin{equation}
    \textbf{m} = ReLU(S)
    \label{eq:gc:relu}
\end{equation}

Finally, as $\textbf{m}$ has a spatial resolution of $[H' \times W']$, which is generally not equal to the image size, Grad-CAM upsamples back to the original spatial resolution, to $[H \times W]$ in \equationref{eq:gc:upsample}

\begin{equation}
    \textbf{m}_{GC} = UP(\textbf{m}).
    \label{eq:gc:upsample}
\end{equation}

\subsection{Guided Grad-CAM}
Guided Grad-CAM can be defined as a hybrid interpretation method that unites Guided Backpropagation and Grad-CAM. This method performs element-wise multiplication between the Guided Backpropagation and Grad-CAM as:

\begin{equation}
    \textbf{m}_{GGC} = \textbf{m}_{GC} \odot \textbf{m}_{GBP}.
\end{equation}

\subsection{NormGrad}

Different from Grad-CAM, NormGrad \cite{Rebuffi20} leverages $\textbf{g}^{out}$ directly without estimating a weight vector from either the gradients or activations. It also makes it possible to use $\Tilde{\textbf{x}}^{out}$, which utilizes us to have the gradient and activations at a particular spot within a neural network. This is caused by using a virtual identity layer $\Tilde{k_t}$, which is present right after the layer of interest $k_t$. In addition, the choice of a virtual identity layer also alters the spatial contribution, as demonstrated in \tableref{tab:ng:sc}.

\begin{table*}[]
  \caption{Spatial contributions, shapes and \textcolor{mycolor}{saliency map formulas} of different virtual identity layer choices.}
    \centering
        \begin{tabular}{|r|c|c|c|}
        \hline
        \textbf{Layer}        & \textbf{Spatial Contribution}                               & \textbf{Shape}        & \textcolor{mycolor}{\textbf{Saliency Map}}            \\ \hline
        Bias         &  $\textbf{g}^{out}_{u}$     & $K'$   &  $ \lVert \textbf{g}^{out} \rVert_F $ \\ \hline
        Scaling      & $\textbf{g}^{out}_{u} \odot \textbf{x}^{out}_{u}$  & $K'$ & $ \lVert \textbf{g}^{out} \odot \textbf{x}^{out} \rVert_F $   \\ \hline
        Conv $N \times N$ & $\textbf{g}^{out}_u  {\textbf{x}^{out}_{u, N \times N}}^\intercal$      & $K' \times N^2K'$ & $ \lVert \textbf{g}^{out} \rVert_F $  $ \lVert \textbf{x}^{out}_{N \times N} \rVert_F $\\ \hline
        \end{tabular}
    \label{tab:ng:sc}
\end{table*}

The first row of \tableref{tab:ng:sc} demonstrates the spatial contribution of a bias layer, which is directly equal to the upstream gradient. As a consequence, the saliency map generated by using the bias virtual identity layer would equal the Frobenius Norm of the upstream gradient of the layer, $\Tilde{k}_t$. 
The second row corresponds to a scaling layer where the spatial contribution is obtained by calculating the element-wise multiplication of the upstream gradient, $\textbf{g}^{out}$, with the activations, $\textbf{x}^{out}$. This results in a saliency map that is directly equal to the Frobenius Norm of the spatial contribution. However, additional operations are needed to be used for calculating the spatial contributions and the corresponding saliency maps for an $N \times N$ convolutional virtual identity layer. Suppose that the output relation of the convolution operation that uses the unfolded version of the input tensor, $\textbf{X}_{N \times N}^{out} \in {\rm I\!R}^{N^2K \times H'W'}$, and the reshaped version of the parameters of the layer $\Tilde{k}_t$, $\Tilde{\textbf{W}}_{N \times N} \in {\rm I\!R}^{K' \times N^2K}$, is expressed in the form of 

\begin{equation}
    \Tilde{\textbf{X}}^{out} = \Tilde{\textbf{W}}_{N \times N} \textbf{X}_{N \times N}^{out},
\end{equation}
which uses matrix multiplication to express $\Tilde{\textbf{X}}^{out} \in {\rm I\!R}^{K' \times H'W'}$. The unfold operation extracts $N \times N$ patches from the input tensor, $\textbf{x}_{out}$ in order to accelerate the convolution operation. If we denote each of the column elements of $\textbf{X}_{N \times N}^{out}$ as $\textbf{x}_{u, N \times N}^{out} \in {\rm I\!R}^{N^2 K'}$, we can find the spatial contribution by calculating the gradient of loss w.r.t. $\Tilde{\textbf{W}}_{N \times N}$ that

\begin{equation}
    \frac{dL}{d\Tilde{\textbf{W}}_{N \times N}} = \sum_{u \in \Omega} \frac{d}{d\Tilde{\textbf{W}}}_{N \times N} \langle \textbf{g}^{out}_u, \Tilde{\textbf{W}}_{N \times N} \textbf{x}^{out}_{u, N \times N} \rangle = \sum_{u \in \Omega} \textbf{g}_u^{out} {\textbf{x}^{out}_{u, N \times N}}^\intercal.
\end{equation}
Since the Frobenius Norm of an outer product can be decomposed into the multiplication of Frobenius Norms, we can calculate the resulting saliency map by first calculating the Frobenius Norms of $\textbf{g}^{out}$ and $\textbf{x}^{out}_{N \times N}$ separately and multiplying these two matrices. The general procedure of transforming the spatial contributions into saliency maps is defined as the aggregation procedure, which is concluded by upsampling the saliency maps into the original image size, $[H, W]$, in order to obtain the final output for a single layer scenario, $\textbf{m}_{NG}$.

A different aspect of NormGrad is its ability to combine the saliency maps of multiple layers within a neural network generated by using gradients and activations. In this paper, we used the uniform setting for combined saliency map calculation which calculates the geometric mean of the provided saliency maps. In addition, for a fair assessment, we use identical virtual identity layers when we would like to obtain a combined saliency map. Hence, this enables us to have a unified and unbiased information resource about what the network has learned by interleaving the output of different layers.  If we have $J$ saliency maps following the aggregation procedure, the combined saliency map $\textbf{M}_{NG}$ is obtained by 

\begin{equation}
    \textbf{M}_{NG} = \Pi_{j=1}^J  \sqrt[J]{\textbf{m}_{NG}^{(j)}},
    \label{eq:combine}
\end{equation}

where $\textbf{M}_{NG}$ becomes the final output as the whole procedure is demonstrated in \figureref{fig:normgrad}.

\section{Experimental Results}
\label{sec:experiments}

We perform our experiments on two datasets which are Object-CXR \cite{CXR20} and LVOT dataset. We provide the details about the datasets in Section \ref{sec:datasets}, our general approach in evaluating the saliency maps in Section \ref{sec:evaluation}, measure the effect of smoothing the saliency maps in Section \ref{sec:smoothing}, make a comparison during initialization and post-training phase in Section \ref{sec:init_vs_train}, and compare the variability across different neural network models in Section \ref{sec:reproducibility}. We make the code and the experiments available on https://github.com/canerozer/explainable-iqa. 

\subsection{Datasets}
\label{sec:datasets}
Object-CXR is a benchmarking dataset, whose objective is to recognize and localise the foreign objects on Chest X-Rays. It contains a total of $10,000$ Chest X-ray images, $5,000$ of which include foreign objects and $5,000$ without.

Left Ventricular Outflow Tract (LVOT) detection is a cardiac MR dataset, where the presence of LVOT is a local quality issue that hinders the accurate analysis of atrial regions. The dataset is composed of a range of 4-chamber cardiac MRI scans from $690$ 2D$+$time patient records with an even number of samples with and without LVOT. We apply a patient-wise splitting on the dataset by using $551$ patients for training, $69$ patients for validation and $70$ patients for testing. Since our network is designed to handle 2D images, we consider each of the slices of the four-chamber view independently. Hence, we expand the number to $8,682$ good quality and $8,522$ LVOT samples. This dataset is constructed by acquiring the images from the Istanbul Mehmet Akif Ersoy Thoracic and Cardiovascular Surgery Training and Research Hospital, where the local ethics committee has waived the need for informed consent for using the anonymised patient records in this retrospective study.

\subsubsection{Chest X-Ray Foreign Object Detection}
We train ResNet-34 \cite{He2016} and EfficientNet-B0 \cite{Tan2019} models, which both predict whether there is at least one foreign object present or not, given a resized $600 \times 600 \times 3$ image. We fine-tune these models, which have been previously trained on the ImageNet dataset \cite{Deng2009}, for $20$ epochs using a batch size of $16$ and a cross-entropy loss function. We adapt our input by duplicating it on the channel axis three times and replace the last layer of the pre-trained model, which now has two output neurons, with random parameters by using He initialisation \cite{He2015}. In order to optimise the parameters during the training procedure, we use Stochastic Gradient Descent with Momentum optimiser, whose learning rate is set to $0.005$ and the learning rate is reduced by a factor of $10$ in every five epochs. Also, to prevent overfitting during training, we use colour jittering, affine transformations, and horizontal flips as data augmentations. Finally, we preserve the best-performing model, depending on the validation accuracy. The performance of the models in terms of AUC score and accuracy is present in \tableref{tab:appendixclsperformance}, where we see that the peak performance has been achieved in the first trial of the ResNet-34 (R34) model by reaching an accuracy of $0.870$ and the AUC score of $0.938$.

\subsubsection{Left Ventricular Outflow Tract (LVOT) Detection}
We train ResNet-50 and EfficientNet-B0 models, which predict whether there exists LVOT or not, given a $224 \times 224 \times 3$ input image. As in the foreign object detection task, we take a ImageNet pre-trained model and fine-tune the model with Stochastic Gradient Descent with Momentum optimiser, whose learning rate is set to $0.0002$ and weight decay to $0.0005$. The fine-tuning procedure takes 60 epochs using a batch size of $64$ when the cross-entropy loss function is used. Also, the same types of data augmentations for the foreign object detection task were used. 
The slice-wise accuracies and AUC scores for each of our trials can be found in \tableref{tab:appendixclsperformance}. Our best LVOT detection model hits a top performance in the first trial of the ResNet-50 (R50) model by reaching an accuracy and AUC score of $0.971$ and $0.998$, respectively.
For the rest of the paper, we name this dataset as LVOT dataset.

\subsection{Experimental Evaluation on the Saliency Maps}
\label{sec:evaluation}
In order to analyse the saliency maps, we present a qualitative and quantitative analysis of the validation and testing splits of the Object-CXR and LVOT datasets. We start with the effect of applying smoothing on the saliency maps generated by these methods. Then, we move on with the randomisation experiments in which we test the saliency map evaluation when (i) model parameters are entirely randomised and (ii) convolutional layer parameters are adopted from ImageNet, and the classifier is randomised. We compare these results with the trained versions of the models as we show their repeated Pointing Game scores. Finally, we demonstrate our reproducibility results which are designed for observing the agreement in the Pointing Game accuracies across different models, namely, ResNet and EfficientNet. 

For highlighting the abilities of NormGrad, we do not only stick to using the penultimate layer (conv4.2), which we name NormGrad Single. We also use the spatial contribution at $J=4$ different layers, e.g., conv2.0, conv3.0, conv4.0 and conv4.2, and call it NormGrad Combined as a result of aggregating spatial contributions for the ResNet architectures. For EfficientNet models instead, despite NormGrad single also using the penultimate layer of the model (features.8.2), we aggregate the spatial contribution at $J=10$ layers, namely, features.0.0, features.1.0, features.2.0, features.3.0, features.4.0, features.5.0, features.6.0, features.7.0, features.8.0, and, features.8.2. Then, we compare our results with other saliency detector baselines such as Grad-CAM \cite{Selvaraju2020}, Guided Grad-CAM \cite{Selvaraju2020}, Guided Backpropagation \cite{Springenberg2015} and Input x Gradient \cite{Shrikumar2016}.

In our study, we use Pointing Game \cite{Zhang16} for quantitatively analysing the abilities of saliency detectors. Our purpose is to detect if saliency maps show a correspondence with the ground-truth bounding boxes, given a medical scan. In Pointing Game, we measure this correspondence by finding the maximum value of a saliency map and then checking the maximum value's proximity to the ground-truth with an offset, $\tau$. We take the default value, $\tau=15$, for both datasets. 
If the maximum value is close enough to the bounding boxes, we define the saliency map to be accurate. If we name the number of accurate saliency maps with $T$ and inaccurate ones with $F$, we can derive an accuracy metric $A$ such that,
\begin{equation}
    A = \frac{T}{T + F}.
\end{equation}

\subsection{Effect of Smoothing the Saliency Maps}
\label{sec:smoothing}
\begin{figure*}
    \centering
    \includegraphics[width=\textwidth]{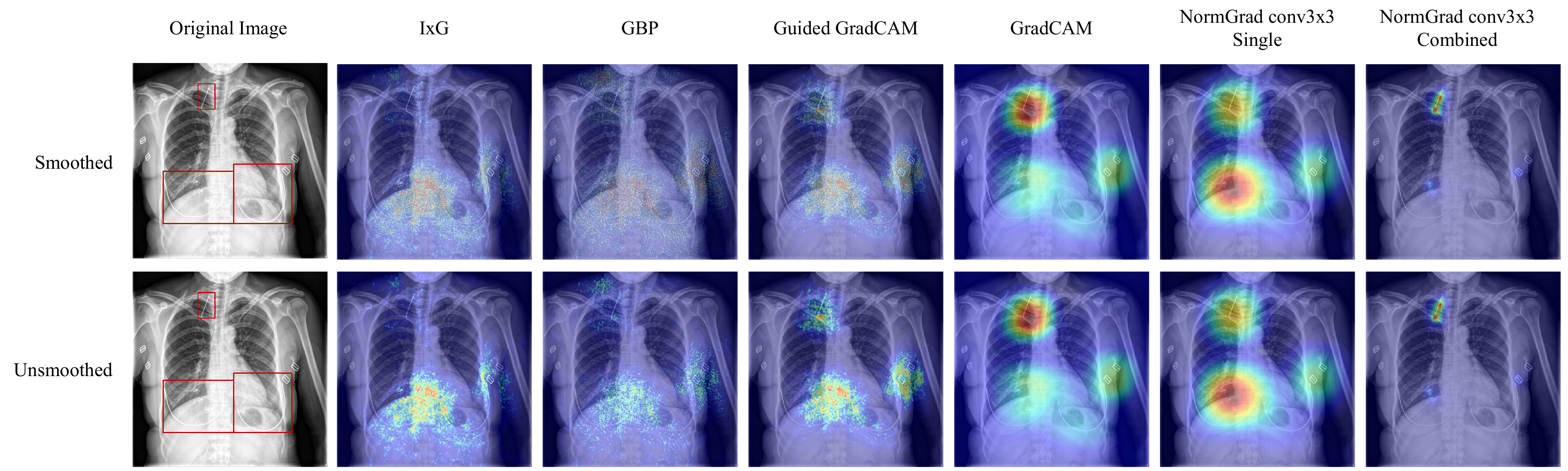}
    \caption{Smoothing operation when used on different interpretability methods. Although IxG (Input x Grad), GBP (Guided Backpropagation) and Guided GradCAM's saliency outputs change heavily, we do not see such drastic changes in the saliency maps generated by GradCAM and NormGrad.}
    \label{fig:smooth_vs_not}
\end{figure*}

This study investigates the effects of smoothing on the performance of various saliency detectors for medical image quality problems. Motivated by the work of \cite{Zhang16}, which suggests that some methods as Grad x Input, tend to output noisy saliency maps and require a smoothing operation to reduce sparsity and noise, a Gaussian kernel with a standard deviation of $1.0$ was applied to smooth the saliency maps. The results of smoothing the saliency maps are shown in \figureref{fig:smooth_vs_not}, where it is observed that saliency maps generated by GradCAM and NormGrad are stable, but Input x Grad (IxG), Guided Backpropagation (GBP) and Guided GradCAM are not. We also see that the conv3x3 combined model of NormGrad successfully focuses on all target foreign objects of interest, especially on the clip at the top. Furthermore, \tableref{tab:lvot_ocxr_smooth_all} demonstrates that the changes in Pointing Game accuracies are insignificant for both LVOT and Object-CXR datasets when NormGrad or GradCAM are used. However, smoothing becomes increasingly crucial for methods that directly use gradient information, as seen in the first three rows of \tableref{tab:lvot_ocxr_smooth_all}. Therefore, smoothing was applied in all comparisons to provide fairness across all saliency detectors while considering the noisiness factor of Input x Grad, Guided Backpropagation and Guided Grad-CAM. Despite this, the success of NormGrad is still evident, while the combined versions appear to be more consistent when their performance on both datasets is considered.

\begin{table*}[]
\centering
\caption{Pointing Game accuracies of all available saliency methods when both smoothing is applied and not. Smoothing has only a slight effect on saliency detectors such as NormGrad and GradCAM, whereas it becomes more essential for noisy saliency detectors such as Input x Grad, Guided Backpropagation, and Guided GradCAM. Best performing methods have been shown with boldcase numbers.}
\resizebox{\columnwidth}{!}{
\begin{tabular}{|l|c|c|cc|cc|}
\hline
                       & \multirow{2}{*}{\begin{tabular}[c]{@{}c@{}}Single\\ Layer\end{tabular}} & \multirow{2}{*}{\begin{tabular}[c]{@{}c@{}}Combined\\ Layer\end{tabular}} & \multicolumn{2}{c|}{LVOT}                                        & \multicolumn{2}{c|}{Object-CXR}                                  \\ \cline{4-7} 
                       &                                                                         &                                                                           & \multicolumn{1}{c|}{Unsmoothed}           & Smoothed             & \multicolumn{1}{c|}{Unsmoothed}           & Smoothed             \\ \hline
Input x Grad           & \checkmark                                                              &                                                                           & \multicolumn{1}{c|}{0.000}                & 0.098                & \multicolumn{1}{c|}{0.162}                & 0.670                \\ \hline
Guided Backpropagation & \checkmark                                                              &                                                                           & \multicolumn{1}{c|}{0.000}                & 0.135                & \multicolumn{1}{c|}{0.132}                & 0.648                \\ \hline
Guided GradCAM         & \checkmark                                                              &                                                                           & \multicolumn{1}{c|}{0.051}                & 0.319                & \multicolumn{1}{c|}{0.256}                & 0.640                \\ \hline
GradCAM                & \checkmark                                                              &                                                                           & \multicolumn{1}{c|}{0.582}                & 0.554                & \multicolumn{1}{c|}{0.572}                & 0.546                \\ \hline
NormGrad Scaling       & \checkmark                                                              &                                                                           & \multicolumn{1}{c|}{0.472}                & 0.468                & \multicolumn{1}{c|}{{\ul \textbf{0.850}}} & {\ul \textbf{0.852}} \\ \hline
NormGrad Scaling       &                                                                         & \checkmark                                                                & \multicolumn{1}{c|}{0.627}                & 0.626                & \multicolumn{1}{c|}{0.840}                & 0.838                \\ \hline
NormGrad Conv1x1       & \checkmark                                                              &                                                                           & \multicolumn{1}{c|}{0.431}                & 0.430                & \multicolumn{1}{c|}{0.848}                & 0.848                \\ \hline
NormGrad Conv1x1       &                                                                         & \checkmark                                                                & \multicolumn{1}{c|}{0.639}                & 0.639                & \multicolumn{1}{c|}{0.840}                 & 0.838                \\ \hline
NormGrad Conv3x3       & \checkmark                                                              &                                                                           & \multicolumn{1}{c|}{0.295}                & 0.292                & \multicolumn{1}{c|}{{\ul \textbf{0.850}}} & 0.850                \\ \hline
NormGrad Conv3x3       &                                                                         & \checkmark                                                                & \multicolumn{1}{c|}{{\ul \textbf{0.640}}} & {\ul \textbf{0.649}} & \multicolumn{1}{c|}{0.846}                & 0.846                \\ \hline
\end{tabular}
}
\label{tab:lvot_ocxr_smooth_all}
\end{table*}

\subsection{Randomisation and Repeatability Experiments}
\label{sec:init_vs_train}
In this part of our study, we aim to demonstrate interpretability by measuring the Pointing Game accuracy of the randomly initiated and trained models. We are influenced by the work of \cite{Arun2021} that claims saliency maps have several shortcomings as saliency detectors are not always robust to randomisation, repeatability, and reproducibility. To measure the effect of randomisation on the saliency maps, we examined two different configurations to initialise the models. First, we use a model of all parameters at random (Fully Randomised, FR) and second, we inherit an ImageNet \cite{Deng2009} model for convolutional layer parameters and randomise only the final fully-connected layer (Semi Randomised, SR). Our purpose for analysing the models during the initialisation stage is to see whether it is possible to catch any significant difference in the interpretability maps after training the neural network. Hence, we assess whether we can use this information as an alternative way for measuring a deep learning model's performance. We also would like to see how much it contributes to the interpretability capability of a deep neural network as a consequence of using ImageNet pre-trained weights for a medical imaging task. During full randomisation, we used He initialisation \cite{He2015} to randomise the weight parameters. While repeating these randomisation experiments three times to have an estimate with high confidence, we also train three models with different seeds to demonstrate the performance as a result of training. We also make a comparison between pre- and post-trained models’ performance. In all our experiments, we report the mean and standard deviation of the Pointing Game accuracy. 

\subsubsection{Results for the LVOT experiments}

\begin{figure}
    \centering
    \includegraphics[width=\columnwidth]{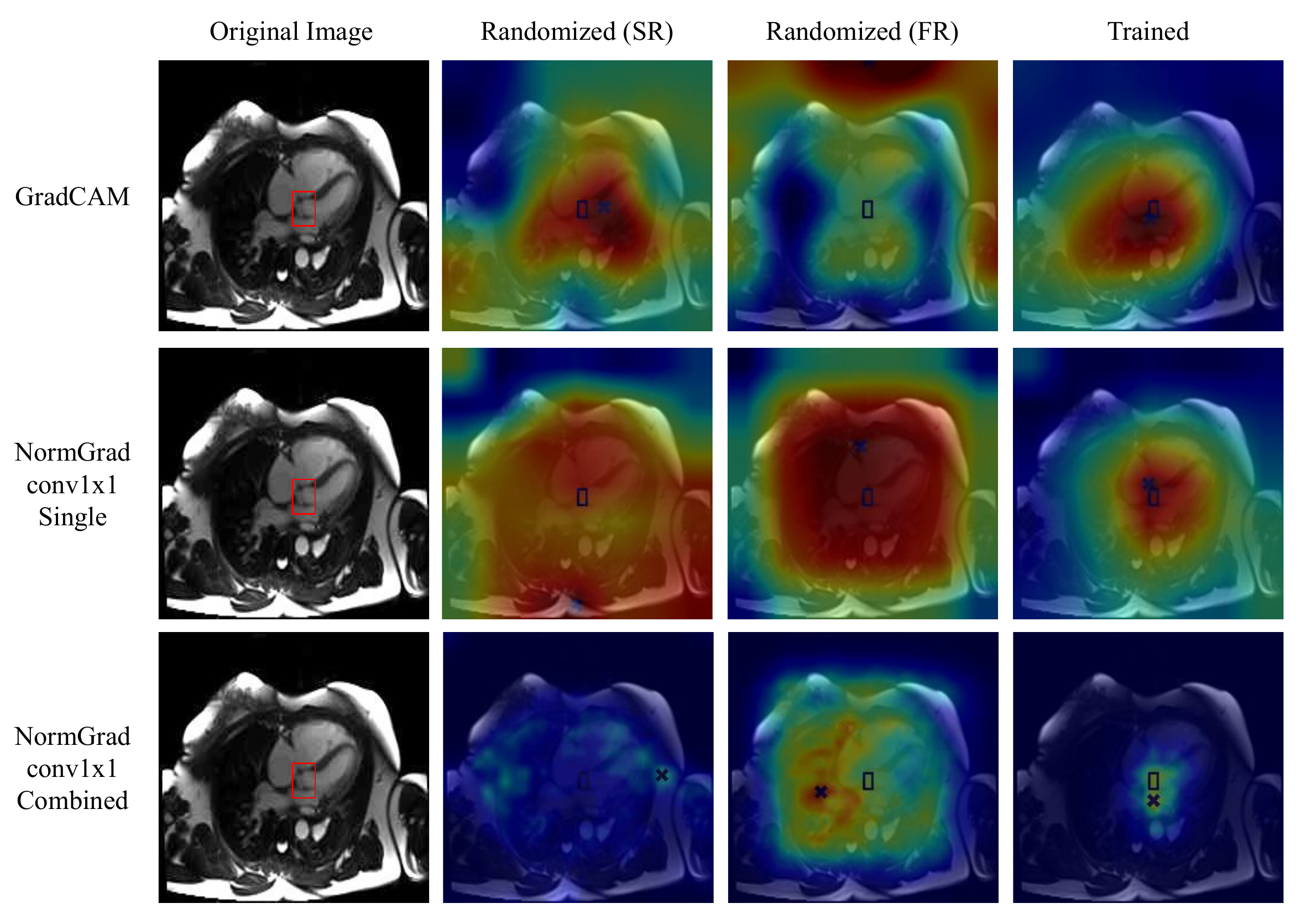}
    \caption{Saliency maps before and after training. Learning appropriate representations has an enormous effect on the quality of saliency maps. Still, combining the saliency map information of different layers is crucial with regard to precision.}
    \label{fig:lvot_random_repeat}
\end{figure}

In \tableref{tab:lvot_random_repeat}, we demonstrate the randomisation and repeatability experiments for the LVOT dataset. First, we notice a significant improvement in the Pointing Game accuracies after training the networks and running all of the saliency detection methods. As we demonstrate in \figureref{fig:lvot_random_repeat}, we associate it with the appropriately-learned representations to fulfil the task and aggregate the information coming from different layers. This also leads to another statement that it is almost obligatory to train the neural networks for making them pointing to relevant regions of interest. 

Secondly, using the combined layer versions of NormGrad can lead to significant improvements in Pointing Game accuracy comparing the performance of single layer saliency detectors. As we switch from GradCAM to the combined layer setting of NormGrad Conv1x1, we are able to improve our Pointing Game accuracy from $0.547$ to $0.611$. We also see further improvements following the adoption of the combined layer setting instead of the single layer setting of NormGrad. The most significant performance improvement can be seen in the NormGrad Conv3x3 setting by reaching a repeated Pointing Game accuracy of $0.602$ from $0.430$ as a result of using combined layers. We owe this to NormGrad Combined's efficiency in localising small target regions of interest since it considers the activation maps and gradients of 4 different layers. Finally, we cannot come up with a conclusion for the LVOT dataset that using pre-trained parameters would help improve the initial performance of the Pointing Game accuracy since, for some of the NormGrad settings, fully-randomised models outweigh their semi-randomised counterparts.  We can relate this to the LVOT region being too small compared to the size of the image, which can only be combated by using the combined configuration of NormGrad that uses gradients and activations of layers at different stages.

\begin{figure*}
    \centering
    \includegraphics[width=\textwidth]{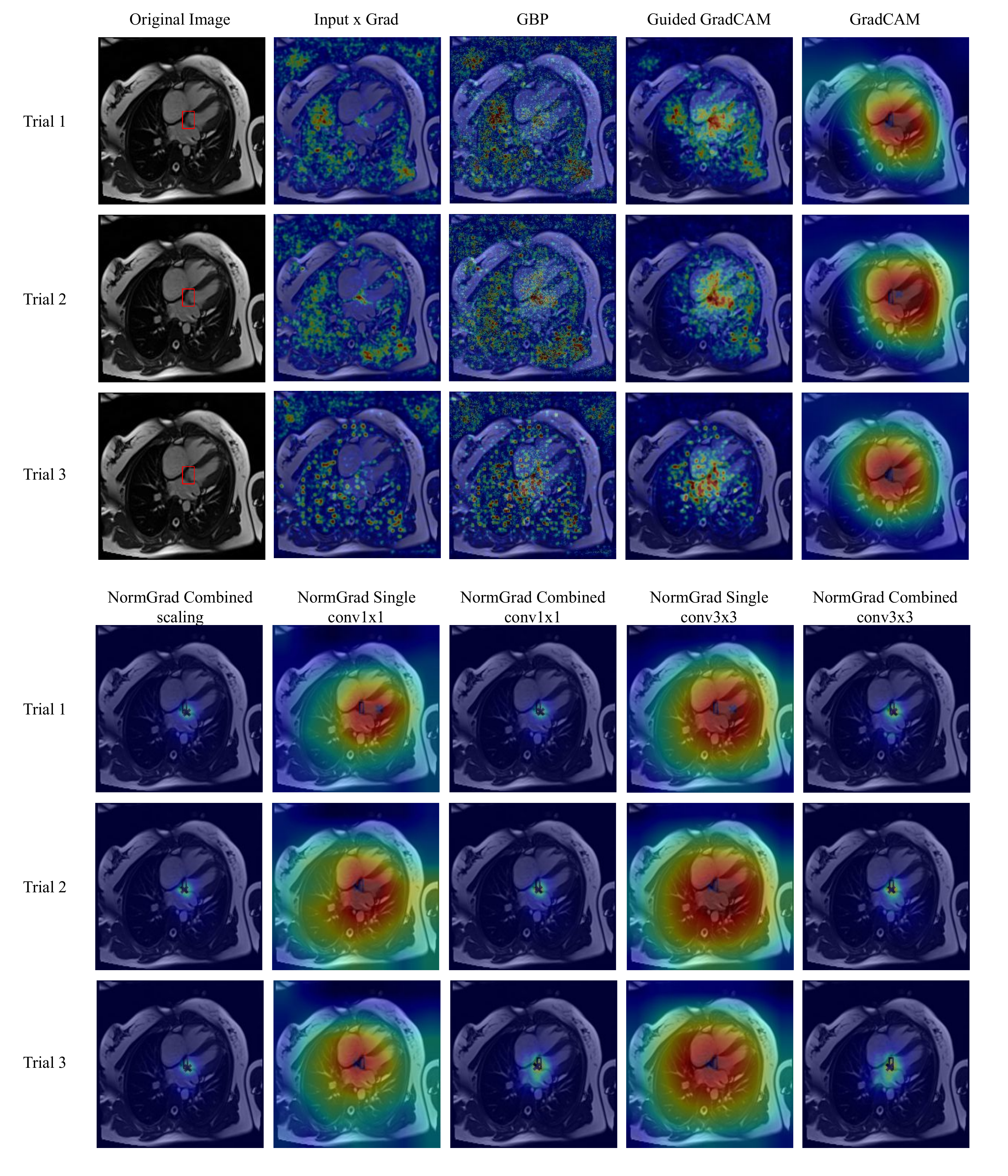}
    \caption{Repeatability results for a sample from LVOT Test Set (Best viewed in zoom). Consistent results are achieved for GradCAM, Guided GradCAM, and NormGrad methods, while the precision is the best for the Combined setting of NormGrad. LVOT region is shown with a bounding box, while the maximum point is shown with a cross. 
    }
    \label{fig:lvot_repeat}
\end{figure*}

The consistency of NormGrad methods after retraining the models is also visually evaluated by comparing it with the baseline methods, where we see noticeable changes in saliency maps. As in \figureref{fig:lvot_repeat}, Input x Grad and Guided Backpropagation (GBP) are problematic due to the additional focus on the background, GradCAM may exhibit some saliency outside the cardiac area or fail to focus on the target region of interest in the event of misclassification, and Guided GradCAM is reliant on the performance of GradCAM and GBP. However, NormGrad satisfies the consistency by covering the cardiac area over saliency maps and even further improves the precision by combining the saliency maps generated by multiple layers.

\begin{table*}[]
\centering
\caption{Pointing Game accuracies of all available saliency methods when the model is initiated via ImageNet parameters except for the final fully-connected layer (Semi randomised, SR), random parameters at full (Fully randomised, FR), and repeatedly trained and examined (Repeated) for the LVOT dataset. Trained models illustrate a superior performance over the randomised models, and NormGrad Combined configurations outweigh the baselines after training.}
\resizebox{\columnwidth}{!}{
\begin{tabular}{|l|c|c|lll|}
\hline
                       & \multirow{2}{*}{\begin{tabular}[c]{@{}c@{}}Single\\ Layer\end{tabular}} & \multirow{2}{*}{\begin{tabular}[c]{@{}c@{}}Combined\\ Layer\end{tabular}} & \multicolumn{3}{c|}{LVOT}                                                                                   \\ \cline{4-6} 
                       &                                                                         &                                                                           & \multicolumn{1}{c|}{SR}              & \multicolumn{1}{c|}{FR}              & \multicolumn{1}{c|}{Repeated} \\ \hline
Input x Grad           & \checkmark                                                              &                                                                           & \multicolumn{1}{l|}{0.001$\pm$0.001} & \multicolumn{1}{l|}{0.048$\pm$0.015} & 0.106$\pm$0.058               \\ \hline
Guided Backpropagation & \checkmark                                                              &                                                                           & \multicolumn{1}{l|}{0.004$\pm$0.002} & \multicolumn{1}{l|}{0.103$\pm$0.041} & 0.152$\pm$0.075               \\ \hline
Guided GradCAM         & \checkmark                                                              &                                                                           & \multicolumn{1}{l|}{0.027$\pm$0.025} & \multicolumn{1}{l|}{0.044$\pm$0.057} & 0.325$\pm$0.075               \\ \hline
GradCAM                & \checkmark                                                              &                                                                           & \multicolumn{1}{l|}{0.061$\pm$0.104} & \multicolumn{1}{l|}{0.008$\pm$0.014} & 0.547$\pm$0.020               \\ \hline
NormGrad Scaling       & \checkmark                                                              &                                                                           & \multicolumn{1}{l|}{0.069$\pm$0.029} & \multicolumn{1}{l|}{0.007$\pm$0.002} & 0.497$\pm$0.140               \\ \hline
NormGrad Scaling       &                                                                         & \checkmark                                                                & \multicolumn{1}{l|}{0.066$\pm$0.038} & \multicolumn{1}{l|}{0.057$\pm$0.023} & 0.600$\pm$0.054               \\ \hline
NormGrad Conv1x1       & \checkmark                                                              &                                                                           & \multicolumn{1}{l|}{0.048$\pm$0.000} & \multicolumn{1}{l|}{0.011$\pm$0.016} & 0.479$\pm$0.139               \\ \hline
NormGrad Conv1x1       &                                                                         & \checkmark                                                                & \multicolumn{1}{l|}{0.054$\pm$0.011} & \multicolumn{1}{l|}{0.082$\pm$0.079} & \textbf{0.611$\pm$0.051}      \\ \hline
NormGrad Conv3x3       & \checkmark                                                              &                                                                           & \multicolumn{1}{l|}{0.073$\pm$0.000} & \multicolumn{1}{l|}{0.215$\pm$0.041} & 0.430$\pm$0.185               \\ \hline
NormGrad Conv3x3       &                                                                         & \checkmark                                                                & \multicolumn{1}{l|}{0.118$\pm$0.006} & \multicolumn{1}{l|}{0.057$\pm$0.023} & 0.602$\pm$0.061               \\ \hline
\end{tabular}
}
\label{tab:lvot_random_repeat}
\end{table*}

\subsubsection{Results for the Object-CXR experiments}

\begin{figure}
    \centering
    \includegraphics[width=\columnwidth]{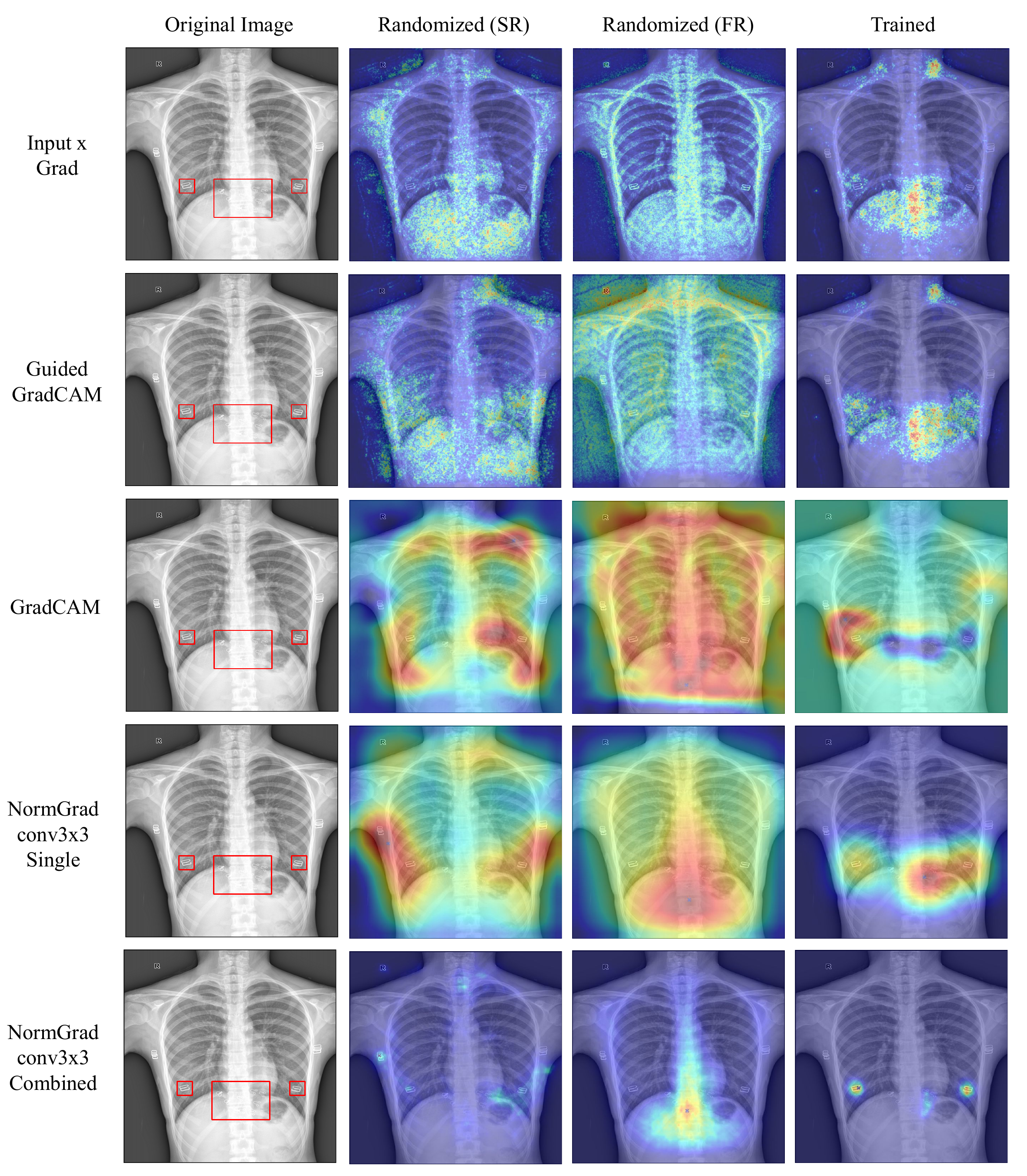}
    \caption{Saliency maps before and after training. Learning appropriate representations has an enormous effect on the quality of saliency maps.}
    \label{fig:ocxr_random_repeat}
\end{figure}

\tableref{tab:ocxr_random_repeat} shows the results for the Object-CXR dataset, where we see a significant improvement in methods after training the models, as in the case of the LVOT dataset. As shown in \figureref{fig:ocxr_random_repeat}, we also visually see this improvement after training the models. However, it is important to note that not all of the trained representations are relevant. As we see in the “Trained” column of \figureref{fig:ocxr_random_repeat}, saliency detectors such as Input x Grad and Guided GradCAM both highlight a region on the left side of the neck, which is not present for NormGrad-based methods, as we see in the bottom two rows of the same figure. The method successfully identifies all of the foreign objects and even improves precision when using the combined setting of NormGrad. In addition, the standard deviations of the repeatability experiments on the Object-CXR dataset of the NormGrad methods are relatively smaller than the baselines' standard deviations, demonstrating the confidence of the NormGrad methods and their utility as an unbiased interpretability measurement tool regardless of repetition in the task. We also see an insignificant difference among the NormGrad methods, and they are all superior to the baseline models. Interestingly, GradCAM has the lowest repeated Pointing Game metric among all of the saliency detectors, which highly questions its ability in terms of pointing to relevant regions through saliency maps, as one instance is shown in \figureref{fig:ocxr_random_repeat}. Overall, NormGrad provides the best results, achieving the highest mean repeated Pointing Game performance of $0.853$ by using the single layer and Conv3x3 setting. Additionally, unlike the results for randomisation in the LVOT detection task, we see an improvement in the Pointing Game accuracy as a result of using ImageNet features. We attribute this improvement to the increased area of the foreign objects of interest.

\begin{figure*}
    \centering
    \includegraphics[width=\textwidth]{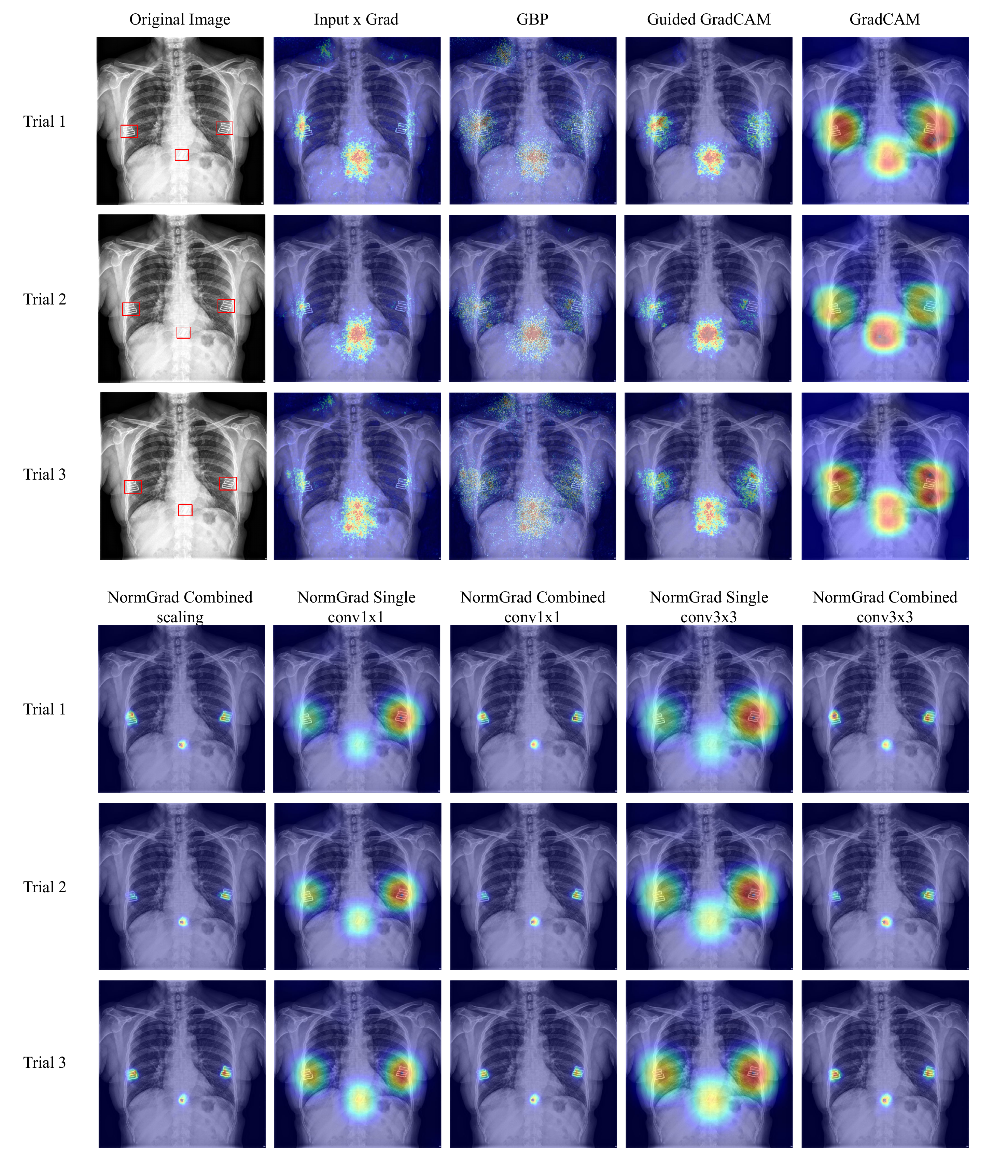}
    \caption{Repeatability results for a sample from Object-CXR Test Set (Best viewed in zoom). NormGrad methods are performing better than the other saliency detection methods in terms of faithfulness.}
    \label{fig:ocxr_repeat}
\end{figure*}

Similarly, as done for the previous dataset, we visually investigate the faithfulness of NormGrad methods, where one case is shown in \figureref{fig:ocxr_repeat}. In this regard, we see that in some of the trials, Input x Grad and Guided GradCAM fail to focus on at least one of the foreign objects, Guided Backpropagation (GBP) may point to the different regions other than the foreign objects of interest, and GradCAM outputs tend to have skewness on the saliency maps. However, all of the NormGrad saliency maps successfully localise most of the foreign objects with a limited number of cases of incorrectly salient regions. Furthermore, the use of combined saliency maps improves the precision by utilising the information that is present on different layers. This makes it more effective to use these saliency maps in potential down-stream tasks.

\begin{table*}[]
\centering
\caption{Pointing Game accuracies of all available saliency methods when the model is initiated via ImageNet parameters except for the final fully-connected layer (Semi randomised, SR), random parameters at full (Fully randomised, FR), and repeatedly trained and examined (Repeated) for the Object-CXR dataset. It is clear that trained models illustrate a significant performance over the randomised models, and all NormGrad configurations outweigh the baselines after training.}
\resizebox{\columnwidth}{!}{
\begin{tabular}{|l|c|c|lll|}
\hline
                       & \multirow{2}{*}{\begin{tabular}[c]{@{}c@{}}Single\\ Layer\end{tabular}} & \multirow{2}{*}{\begin{tabular}[c]{@{}c@{}}Combined\\ Layer\end{tabular}} & \multicolumn{3}{c|}{Object-CXR}                                                                             \\ \cline{4-6} 
                       &                                                                         &                                                                           & \multicolumn{1}{c|}{SR}              & \multicolumn{1}{c|}{FR}              & \multicolumn{1}{c|}{Repeated} \\ \hline
Input x Grad           & \checkmark                                                              &                                                                           & \multicolumn{1}{l|}{0.111$\pm$0.028} & \multicolumn{1}{l|}{0.185$\pm$0.041} & 0.663$\pm$0.013               \\ \hline
Guided Backpropagation & \checkmark                                                              &                                                                           & \multicolumn{1}{l|}{0.091$\pm$0.008} & \multicolumn{1}{l|}{0.160$\pm$0.007} & 0.640$\pm$0.053               \\ \hline
Guided GradCAM         & \checkmark                                                              &                                                                           & \multicolumn{1}{l|}{0.166$\pm$0.055} & \multicolumn{1}{l|}{0.200$\pm$0.028} & 0.645$\pm$0.012               \\ \hline
GradCAM                & \checkmark                                                              &                                                                           & \multicolumn{1}{l|}{0.200$\pm$0.066} & \multicolumn{1}{l|}{0.093$\pm$0.103} & 0.545$\pm$0.018               \\ \hline
NormGrad Scaling       & \checkmark                                                              &                                                                           & \multicolumn{1}{l|}{0.279$\pm$0.021} & \multicolumn{1}{l|}{0.126$\pm$0.032} & 0.852$\pm$0.000               \\ \hline
NormGrad Scaling       &                                                                         & \checkmark                                                                & \multicolumn{1}{l|}{0.320$\pm$0.029} & \multicolumn{1}{l|}{0.147$\pm$0.053} & 0.843$\pm$0.005               \\ \hline
NormGrad Conv1x1       & \checkmark                                                              &                                                                           & \multicolumn{1}{l|}{0.280$\pm$0.000} & \multicolumn{1}{l|}{0.153$\pm$0.025} & 0.851$\pm$0.006               \\ \hline
NormGrad Conv1x1       &                                                                         & \checkmark                                                                & \multicolumn{1}{l|}{0.325$\pm$0.004} & \multicolumn{1}{l|}{0.185$\pm$0.027} & 0.839$\pm$0.010               \\ \hline
NormGrad Conv3x3       & \checkmark                                                              &                                                                           & \multicolumn{1}{l|}{0.298$\pm$0.000} & \multicolumn{1}{l|}{0.153$\pm$0.017} & \textbf{0.853$\pm$0.003}      \\ \hline
NormGrad Conv3x3       &                                                                         & \checkmark                                                                & \multicolumn{1}{l|}{0.334$\pm$0.016} & \multicolumn{1}{l|}{0.193$\pm$0.027} & 0.845$\pm$0.005               \\ \hline
\end{tabular}
}
\label{tab:ocxr_random_repeat}
\end{table*}

\subsection{Reproducibility Experiments}
\label{sec:reproducibility}
Another criterion for evaluating the faithfulness of saliency detectors is their reproducibility, where we calculate their performance whenever we train a completely different network architecture. On this subject, we compare the saliency detection performance of ResNet with EfficientNet-B0 \cite{Tan2019} for both of the tasks, and propose a basic metric named Difference of Means (DoM) for measuring the consistency under a change in network architecture. To calculate the DoM measure, we obtain the repeated Pointing Game metric for each of the network architectures, and we simply the absolute difference between the means of it. To maintain consistency across different models, we would like to have a lower DoM score reaching down to 0. 

\begin{table*}[]
\centering
\caption{Pointing Game performances on two different network architectures: ResNet (R50/R34) vs EfficientNet (EB0). The minimum Difference of Means (DoM) score is obtained in the conv3x3 combined layer setting of NormGrad for the LVOT dataset and conv1x1 single layer setting of NormGrad for the Object-CXR dataset. Also, mind the difference between the DoM score of the baseline methods and the combined layer setting of NormGrad.}
\resizebox{\linewidth}{!}{
\begin{tabular}{|l|c|c|ccc|ccc|}
\hline
                       & \multirow{2}{*}{\begin{tabular}[c]{@{}c@{}}Single\\ Layer\end{tabular}} & \multirow{2}{*}{\begin{tabular}[c]{@{}c@{}}Combined\\ Layer\end{tabular}} & \multicolumn{3}{c|}{LVOT}                                                                                            & \multicolumn{3}{c|}{Object-CXR}                                                                                      \\ \cline{4-9} 
                       &                                                                         &                                                                           & \multicolumn{1}{c|}{R50}                      & \multicolumn{1}{c|}{EB0}                      & DoM                  & \multicolumn{1}{c|}{R34}                      & \multicolumn{1}{c|}{EB0}                      & DoM                  \\ \hline
Input x Grad           & \checkmark                                                              &                                                                           & \multicolumn{1}{c|}{0.106$\pm$0.058}          & \multicolumn{1}{c|}{0.218$\pm$0.042}          & 0.112                & \multicolumn{1}{c|}{0.663$\pm$0.013}          & \multicolumn{1}{c|}{0.617$\pm$0.036}          & 0.046                \\ \hline
Guided Backpropagation & \checkmark                                                              &                                                                           & \multicolumn{1}{c|}{0.152$\pm$0.075}          & \multicolumn{1}{c|}{0.318$\pm$0.049}          & 0.166                & \multicolumn{1}{c|}{0.640$\pm$0.053}          & \multicolumn{1}{c|}{0.595$\pm$0.045}          & 0.045                \\ \hline
Guided GradCAM         & \checkmark                                                              &                                                                           & \multicolumn{1}{c|}{0.325$\pm$0.075}          & \multicolumn{1}{c|}{0.453$\pm$0.035}          & 0.128                & \multicolumn{1}{c|}{0.645$\pm$0.012}          & \multicolumn{1}{c|}{0.784$\pm$0.005}          & 0.139                \\ \hline
GradCAM                & \checkmark                                                              &                                                                           & \multicolumn{1}{c|}{0.547$\pm$0.020}          & \multicolumn{1}{c|}{0.569$\pm$0.027}          & 0.022                & \multicolumn{1}{c|}{0.545$\pm$0.020}          & \multicolumn{1}{c|}{0.771$\pm$0.005}          & 0.226                \\ \hline
NormGrad Scaling       & \checkmark                                                              &                                                                           & \multicolumn{1}{c|}{0.497$\pm$0.140}          & \multicolumn{1}{c|}{\textbf{0.778$\pm$0.020}} & 0.281                & \multicolumn{1}{c|}{0.852$\pm$0.000}          & \multicolumn{1}{c|}{0.857$\pm$0.010}          & 0.005                \\ \hline
NormGrad Scaling       &                                                                         & \checkmark                                                                & \multicolumn{1}{c|}{0.600$\pm$0.054}          & \multicolumn{1}{c|}{0.620$\pm$0.010}          & 0.020                & \multicolumn{1}{c|}{0.843$\pm$0.005}          & \multicolumn{1}{c|}{0.856$\pm$0.002}          & 0.013                \\ \hline
NormGrad Conv1x1       & \checkmark                                                              &                                                                           & \multicolumn{1}{c|}{0.479$\pm$0.139}          & \multicolumn{1}{c|}{0.754$\pm$0.041}          & 0.275                & \multicolumn{1}{c|}{0.851$\pm$0.006}          & \multicolumn{1}{c|}{0.850$\pm$0.000}          & {\ul \textbf{0.001}} \\ \hline
NormGrad Conv1x1       &                                                                         & \checkmark                                                                & \multicolumn{1}{c|}{\textbf{0.611$\pm$0.051}} & \multicolumn{1}{c|}{0.634$\pm$0.018}          & 0.023                & \multicolumn{1}{c|}{0.839$\pm$0.010}          & \multicolumn{1}{c|}{0.851$\pm$0.008}          & 0.011                \\ \hline
NormGrad Conv3x3       & \checkmark                                                              &                                                                           & \multicolumn{1}{c|}{0.430$\pm$0.185}          & \multicolumn{1}{c|}{0.728$\pm$0.033}          & 0.298                & \multicolumn{1}{c|}{\textbf{0.853$\pm$0.003}} & \multicolumn{1}{c|}{\textbf{0.863$\pm$0.002}} & 0.009                \\ \hline
NormGrad Conv3x3       &                                                                         & \checkmark                                                                & \multicolumn{1}{c|}{0.602$\pm$0.061}          & \multicolumn{1}{c|}{0.607$\pm$0.020}          & {\ul \textbf{0.005}} & \multicolumn{1}{c|}{0.845$\pm$0.005}          & \multicolumn{1}{c|}{0.859$\pm$0.007}          & 0.014                \\ \hline
\end{tabular}
}
\label{tab:reprod}
\end{table*}

\begin{figure*}
    \centering
    \includegraphics[width=\textwidth]{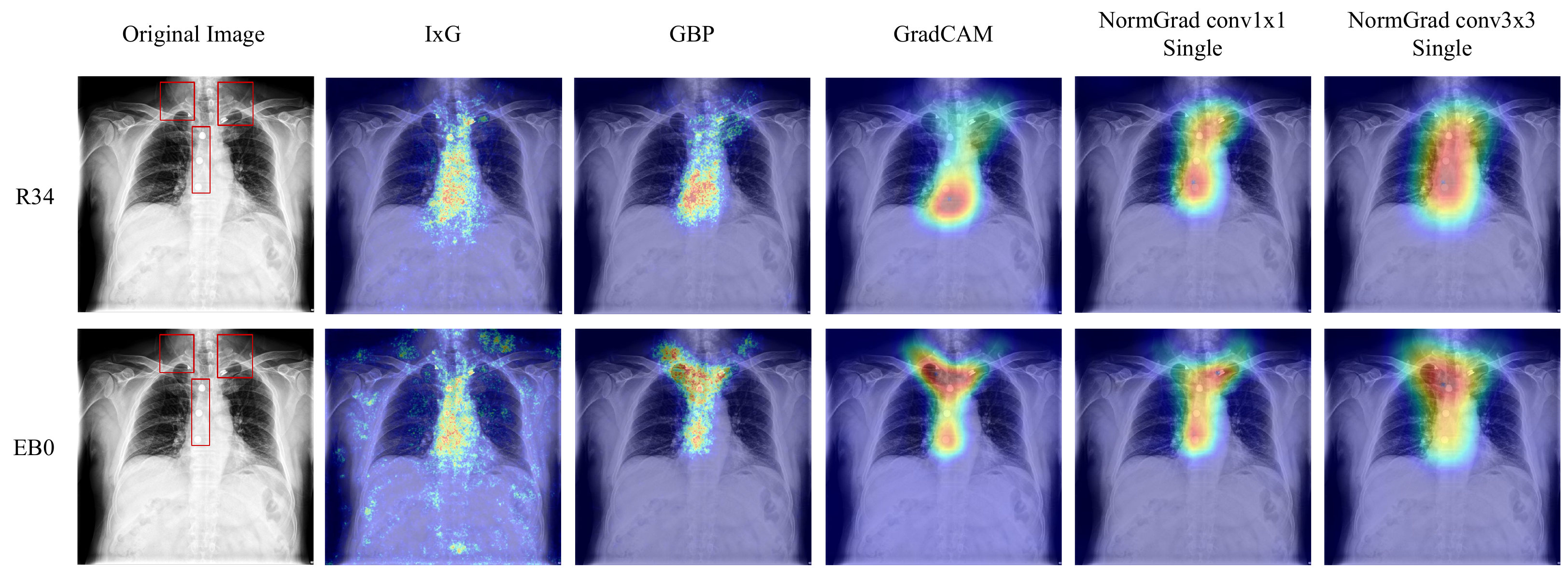}
    \caption{Reproducibility results for a sample from Object-CXR Test Set when ResNet-34 (R34) and EfficientNet-B0 (EB0) architectures are utilized for foreign object detection task. Consistent results are achieved for NormGrad methods since it focuses on all three foreign objects of interest.}
    \label{fig:reprod}
\end{figure*}

In \tableref{tab:reprod}, we observe that the performance of NormGrad methods surpasses the performance of the baselines on both datasets, except for GradCAM when ResNet-50 (R50) architecture was used for the LVOT detection task. Additionally, the best performance is achieved in the single-layer version of NormGrad for the LVOT detection task with the EfficientNet-B0 model, whereas there is a slight offset between the same type of model and the best model for the Object-CXR benchmark. However, we also notice considerable differences between the mean Pointing Game performance when assessing these two types of models under the single-layer version of NormGrad, especially for the LVOT detection task. While such a problem does not exist for the Object-CXR benchmark, we see that the use of combined layers is essential for faithful interpretation of the models, considering the minute DoM measure on both datasets. The differences in the generated saliency maps of different models are reflected in \figureref{fig:reprod} for the Object-CXR dataset.  

\section{Discussion and Conclusions}
\label{sec:diss_conc}

In this work, we proposed an automatic and explainable framework for medical image quality assessment using NormGrad to explain its decisions by pointing to relevant regions of interest. We compared the results of NormGrad with a variety of baseline-category saliency detectors under certain conditions, using the Pointing Game metric. In our study, we first reported the saliency detector performance under smoothing conditions, applying Gaussian smoothing to the produced saliency maps. Then, we assessed the saliency detector performance under different randomization schemes, comparing them with their trained counterparts. Finally, we reported the saliency detector performance when different neural network architectures were used, and proposed a basic but intuitive metric named Difference of Means (DoM) which measures the consistency between the saliency detection performance on two different architectures. 

Comparing other saliency detection methods, we notice that smoothing had a negligible effect on NormGrad, as evidenced by its barely-changed performance on both datasets. Additionally, the method successfully demonstrated the learning capability of a neural network by showing a significant difference in Pointing Game metric between the randomised and repeated versions of a neural network. It is also important to note that, using the combined setting of NormGrad provides a minimum guarantee for providing interpretability from a more global point of view, by integrating the information coming from different layers. While we achieved a significant difference as a result of using the combined version for the LVOT detection task, this difference became insignificant for the Object-CXR task, though we still saw numerous cases where using the combined setting of NormGrad was superior in terms of precision. Finally, we can say that the combined setting of NormGrad is more consistent in terms of the DoM metric and outperforms the baseline models in terms of the Pointing Game score. For this reason, it is evident that NormGrad is a general purpose method that can be used for medical image quality assessment tasks.

Our approach still has limitations. First, it is unlikely to use NormGrad for sensitive predictive tasks such as segmentation in radiation therapy planning \cite{Samarasinghe2021}, as the tissue may vary in terms of size and, in the case of the tissue being sufficiently large comparing the image size, NormGrad may not accurately segment the tissue and cause the radiation therapy damage healthy cells. This example can also be extended to the tasks where there is a sufficiently large amount of data available with pixel-level ground truth annotations, such that it becomes plausible to train a segmentation framework instead. Secondly, the Pointing Game method finds it sufficient for contact with any of the bounding boxes, which makes the objective trivial for images with many target region of interests. Although this does not generate a problem for the LVOT detection task, since there is only a single region of interest, this is slightly problematic for the foreign object detection task, as there are multiple regions of interest in some of the images. 

We consider that our method can be used right after the image acquisition stage for immediately finding the cases with quality errors. By assisting the radiologists by directly pointing to the region of interests through saliency maps, it becomes possible to make an immediate decision for re-acquisition without any human intervention. As for the future work, we plan to involve NormGrad saliency maps during the training stage of the image quality assessment and enhancement frameworks.

\section*{Acknowledgments}
We thank Mehmet Ozan Unal for his contribution to the development of the Object-CXR pre-trained model and Tolga Ok for the fruitful discussions. This paper has been produced benefiting from the 2232 International Fellowship for Outstanding Researchers Program of TUBITAK (Project No: 118C353). However, the entire responsibility of the publication/paper belongs to the owner of the paper. The financial support received from TUBITAK does not mean that the content of the publication is approved in a scientific sense by or TUBITAK.

The paper also benefited from Istanbul Technical University Scientific Research Projects (ITU BAP) funds, grant number 44250.

\bibliographystyle{unsrt}  
\bibliography{references}  

\section*{Appendix}
\subsection{Classification Model Performance}

We report the accuracy and AUC scores for LVOT and Object-CXR datasets for ResNet (R50) and EfficientNet (EB0) architectures. T1, T2, and T3 correspond to our three trials. For both datasets, the best performance is achieved in our first trial, T1. 

\begin{table}[h!]
\centering
\caption{Accuracy and Area Under Curve (AUC) scores for all models.}
\resizebox{0.65\columnwidth}{!}{
\begin{tabular}{|l|cccc|cccc|}
\hline
        & \multicolumn{4}{c|}{LVOT}                                                                          & \multicolumn{4}{c|}{Object-CXR}                                                                    \\ \hline
        & \multicolumn{2}{c|}{R50}                                   & \multicolumn{2}{c|}{EB0}              & \multicolumn{2}{c|}{R34}                                   & \multicolumn{2}{c|}{EB0}              \\ \hline
        & \multicolumn{1}{c|}{Acc} & \multicolumn{1}{c|}{AUC}   & \multicolumn{1}{c|}{Acc} & AUC   & \multicolumn{1}{c|}{Acc} & \multicolumn{1}{c|}{AUC}   & \multicolumn{1}{c|}{Acc} & AUC   \\ \hline
T1 & \multicolumn{1}{c|}{\textbf{0.971}}    & \multicolumn{1}{c|}{\textbf{0.998}} & \multicolumn{1}{c|}{\textbf{0.929}}    & \textbf{0.978} & \multicolumn{1}{c|}{\textbf{0.870}}    & \multicolumn{1}{c|}{\textbf{0.938}} & \multicolumn{1}{c|}{\textbf{0.858}}    & \textbf{0.937} \\ \hline
T2 & \multicolumn{1}{c|}{0.949}    & \multicolumn{1}{c|}{0.994} & \multicolumn{1}{c|}{0.867}    & 0.949 & \multicolumn{1}{c|}{0.853}    & \multicolumn{1}{c|}{\textbf{0.938}} & \multicolumn{1}{c|}{0.857}    & 0.936 \\ \hline
T3 & \multicolumn{1}{c|}{0.943}    & \multicolumn{1}{c|}{0.992} & \multicolumn{1}{c|}{0.915}    & 0.971 & \multicolumn{1}{c|}{0.865}    & \multicolumn{1}{c|}{0.931} & \multicolumn{1}{c|}{0.856}    & 0.936 \\ \hline
\end{tabular}
}
\label{tab:appendixclsperformance}
\end{table}

\end{document}